\documentclass[aps,pra,twocolumn,showpacs,superscriptaddress,a4paper]{revtex4-1}

\usepackage{mathrsfs}
\usepackage{graphicx}

\begin{document}

\title{Ultrafast Molecular Imaging by Laser Induced Electron Diffraction}

\author{M. Peters}
\affiliation{Universit\'e Paris-Sud, Institut des Sciences Mol\'eculaires d'Orsay (CNRS), F-91405 Orsay, France}
\affiliation{D\'epartement de Chimie, Universit\'e Laval, Qu\'ebec, Qu\'ebec, Canada G1K 7P4}

\author{T. T. Nguyen-Dang}
\affiliation{D\'epartement de Chimie, Universit\'e Laval, Qu\'ebec, Qu\'ebec, Canada G1K 7P4}

\author{C. Cornaggia}
\affiliation{CEA IRAMIS, SPAM, Saclay, B\^atiment 522, F-91 191 Gif-sur-Yvette, France}

\author{S. Saugout}
\affiliation{Universit\'e Paris-Sud, Institut des Sciences Mol\'eculaires d'Orsay (CNRS), F-91405 Orsay, France}

\author{E. Charron}
\affiliation{Universit\'e Paris-Sud, Institut des Sciences Mol\'eculaires d'Orsay (CNRS), F-91405 Orsay, France}

\author{A. Keller}
\affiliation{Universit\'e Paris-Sud, Institut des Sciences Mol\'eculaires d'Orsay (CNRS), F-91405 Orsay, France}

\author{O. Atabek}
\affiliation{Universit\'e Paris-Sud, Institut des Sciences Mol\'eculaires d'Orsay (CNRS), F-91405 Orsay, France}

\begin{abstract}

We address the feasibility of imaging geometric and orbital structure of a polyatomic molecule on an attosecond time-scale using the laser induced electron diffraction (LIED) technique. We present numerical results for the highest molecular orbitals of the CO$_2$ molecule excited by a near infrared few-cycle laser pulse. The molecular geometry (bond-lengths) is determined within 3\% of accuracy from a diffraction pattern which also reflects the nodal properties of the initial molecular orbital. Robustness of the structure determination is discussed with respect to vibrational and rotational motions with a complete interpretation of the laser-induced mechanisms.

\end{abstract}

\maketitle

With the development of attosecond laser sources \cite{RevModPhys.72.545}, ultrafast molecular imaging has become a major research topic in modern physics. On one hand, attosecond laser pulses are directly used to image dynamical processes in schemes such as the attosecond pump-probe spectroscopic mapping of molecular motions \cite{PhysRevLett.103.123005} or the interferometric real-time observation of electronic motions \cite{Mauritsson}. On the other hand, a number of schemes have been proposed to image molecular structure, which are all based on the rescattering mechanism \cite{rescattering}. This phenomenon consists of the tunnel ionization of an electron followed by its acceleration and its return by the field, to end with its recollision with the molecular ionic core. The possible outcomes of this mechanism are all relevant to molecular imaging. The elastic scattering of the returning electron with the ion core defines the LIED \cite{LIED,Spanner}, which can be compared with ultrafast electron diffraction using an external electron source \cite{ediffrac}. Inelastic scattering of the returning electron may be accompanied by electronic excitation of the parent ion, its further ionization or the emission of high-energy radiation. This last process, high-harmonic generation (HHG), has been used directly as a probe or indirectly to unveil orbital and molecular structure, and even to image rotational and vibrational motions \cite{smirnovaNat460}.

We wish to assess the feasibility of imaging polyatomic molecular structure, e.g., measuring its geometrical parameters, such as bond-lengths and/or bond-angles), on an attosecond time-scale using the LIED technique. We show how a diffraction pattern constructed from measurable photo-electron momentum distribution encodes informations on the electronic orbital nodal properties but also on the geometry of the nuclei. We assess the robustness of this two-fold structure determination procedure with respect to inevitable uncertainties about the alignment of the molecule relative to the field polarization direction, its bond lengths and angles. We consider a relatively simple model of the CO$_2$ molecule, the choice of which being motivated by different considerations: Being linear in its symmetric equilibrium geometry, it is simple enough, in particular with respect to its alignment properties, and yet with three nuclei, it has enough internal degrees of freedom to make its molecular structure determination challenging.

To concentrate on the problem of reading geometrical informations out of the photo-electron diffraction pattern in momentum space, we develop a theoretical model based on an effective single active electron (SAE) Hamiltonian. Nuclear dynamics and its coupling to the electron ionized out of a definite orbital (initial state) can be ignored at the time scale of a few-cycle pulse. Nuclear degrees of freedom are thus taken into account only through statistical distributions of CO bond-lengths $R$ and polar angles $\theta$ with respect to the laboratory $y$-axis along which CO$_2$ is supposed initially aligned. The intense laser pulse, which monitors the electron dynamics, is linearly polarized along the laboratory $x$-axis and the essential of the electron diffraction process takes place in the $(x,y)$-plane. Electronic wave packets are generated by solving the time-dependent Schr\"odinger equation (atomic units are used):
\begin{equation}
\dot\imath\frac{\partial}{\partial t} \Phi(\vec r, t) = \hat{H}(\vec r, t)\Phi(\vec r, t)
\end{equation}
with a SAE Hamiltonian using a soft-Coulomb potential and an electron-laser dipole coupling in the length gauge:
\begin{equation}
\hat{H}(\vec r, t) = \frac{\vec p^{\,2}}{2}
                   + \sum_{\alpha = 1}^{3} \frac{- Z_{\alpha}(\vec r)}{\sqrt{|\vec r - \vec \rho_{\alpha}|^2 + a_{\alpha}^2}}
                   + \vec r \cdot \vec{\mathscr{E}}(t)\,,
\end{equation}
$\vec r$ and $\vec p\,$ being the electron position and momentum, and $\alpha$ labeling the three nuclei located at fixed position vectors $\vec \rho_{\alpha}$. The laser field $\vec{\mathscr{E}}(t)$ is characterized by the angular frequency $\omega_L$ and the amplitude $\mathscr{E}_0$. The $\vec r$-dependent effective charge $Z_{\alpha}(\vec r) = Z_{\alpha}^\infty + ( Z_{\alpha}^0 - Z_{\alpha}^\infty ) \exp(-|\vec r - \vec \rho_{\alpha}|^2 / \sigma_\alpha^2)$ involves a single adjustable parameter $\sigma_\alpha$. It is defined, together with the softening parameter $a_{\alpha}$, so as to give field-free eigenstates of the SAE Hamiltonian identifiable with the highest occupied molecular orbitals, HOMO$-n$ $(n = 0, 1,2)$, in the sense of giving the same ionization energy and the same nodal properties as the actual molecular orbital of the 22-electron molecule. Numerical wave packet propagation uses a third-order split operator formula on a finite 2D grid continued analytically by projection onto Volkov states for $|\vec{r}\,| > 45\,$\AA.

\begin{figure}[t!]
\begin{center}
\includegraphics[width=0.48\textwidth]{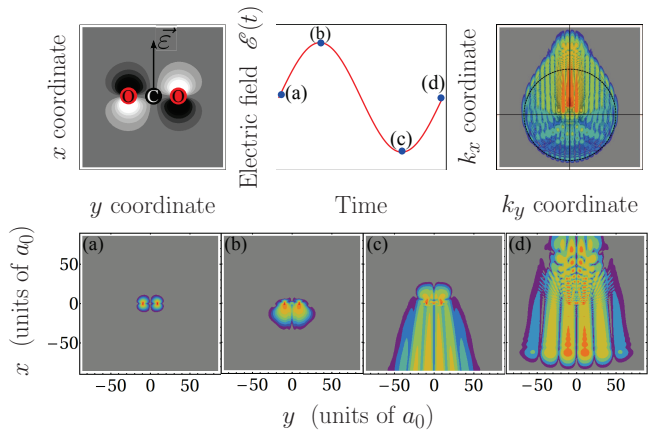}
\caption{\label{figure1} (Color online) Ionized electron motion in laser-driven CO$_2$ molecule. First row, from left to right: Field-free $(\pi_g)$ HOMO of CO$_2$, external field's waveform, and calculated 2D asymptotic electron momentum distribution for the CO bond length 4.8\,\AA. Various snapshots of the calculated 2D electron density are shown in panels (a) to (d) in the second row, corresponding to different times within the optical cycle ($\omega_L = 0.06\,$a.u., $\mathscr{E}_0 = 0.15\,$a.u.) as marked on the field waveform above.}
\end{center}
\end{figure}

The four panels on the second row of Fig.\,\ref{figure1} show the results of the time-propagation from an initial state $\Phi(\vec r, t_0)$ taken as the HOMO, characterized by a perpendicular nodal plane which is preserved through time-evolution, when the laser polarization is orthogonal to the molecular axis. Each panel depicts the electronic density at a different time within a single optical cycle, as indicated by the corresponding letter on the waveform shown in the middle panel of the first row. The most important observations from these spatial wave functions are first the fringes of panel (c) obtained during the forward ($x < 0$) electron motion as a signature of the two oxygen atoms acting as electron ejection sources, and second the rich pattern of interference fringes of panel (d) in the backward ($x > 0$) motion resulting from the electron diffraction analogous to Young's slit experiment with photons.

More relevant for structure determination is the diffraction pattern in reciprocal momentum space, accessible experimentally through time-of-flight electron velocity mapping, and determined numerically by the Fourier transform $\tilde\Phi(k_x,k_y)$ of the asymptotic electron wave packet. To illustrate this, we show, in the rightmost panel on the first row of Fig.\,\ref{figure1}, $|\tilde\Phi(k_x,k_y)|^2$ for the CO bond length $R = 4.8\,$\AA, corresponding to a symmetrically stretched molecule. The features of this momentum distribution, with a succession of vertical interference fringes, are in agreement with the interpretation of Ref.\,\cite{Spanner}. From this distribution, a $k_x$-averaged diffraction pattern can be defined as
\begin{equation}
S(k_y) = \int_\Gamma |\tilde\Phi(k_x,k_y)|^2 \ dk_x\,,
\label{Eq:S}
\end{equation}
where $\Gamma$ denotes the domain of integration. In analogy with Young's double slit experiment, we expect a good resolution of the interference signal if the electron de Broglie wavelength $\lambda_{\mathrm{DB}} = h / p$ is smaller than the slit separation $d = 2R$. This criterion corresponds to $|k| > \pi/R$ and therefore to the high energy range of the photo-electron spectrum. To insure a resolution of the order of 0.5\,\AA, we define here the domain of integration $\Gamma$ as the region where $|k| > 3.15\,$a.u.

\begin{figure}[t!]
\begin{center}
\includegraphics[width=0.48\textwidth]{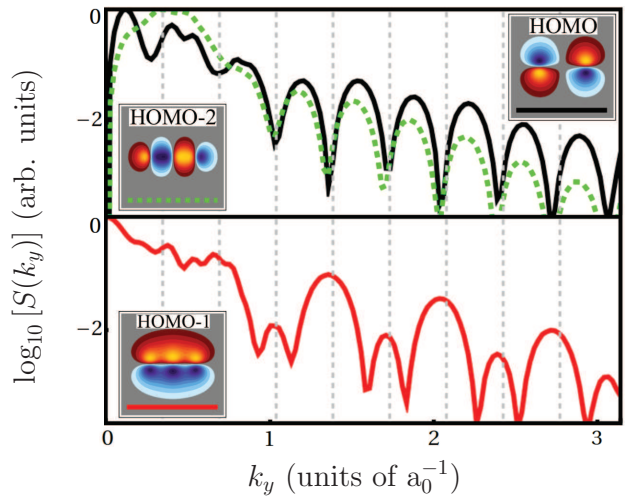}
\caption{\label{figure2} (Color online) Averaged normalized electron diffraction pattern $S(k_y)$ obtained with a single cycle pulse (see Fig.\,\ref{figure1}) for a perfectly aligned molecule, with $R = 4.8\,$\AA. In the upper panel, the black solid and green dashed lines correspond to the diffraction patterns obtained from the $(\pi_g)$ HOMO and $(\sigma_u)$ HOMO-2 molecular orbitals of CO$_2$. The red solid line in the lower panel is for the $(\pi_u)$ HOMO-1 initial state.}
\end{center}
\end{figure}

For a single-cycle pulse, and a perfect alignment situation, the averaged distribution $S(k_y)$ associated with the diffraction of an electron ejected from the HOMO is shown as a black solid line in the upper panel of Fig.\,\ref{figure2}. One can notice that $S(k_y)$ exhibits clear interference fringes at large $k_y$, characterized by a regular succession of peaks, from which geometrical parameters will be inferred. Indeed, that $S(k_y)$ directly gives informations on the nuclear geometry can be seen by the following simple symmetry considerations: The CO$_2$ HOMO initial molecular orbital is anti-symmetric with respect to the perpendicular $\sigma_h$ mirror plane containing the carbon atom, and so is the electronic wave function at all time, assuming perfect alignment of the molecule orthogonal to the laser polarization. It can approximately be written as
\begin{equation}
\Phi_{\pi_g}(\vec r, t ) \propto f(x,y-R,t) - f(x,y+R,t)\,,
\label{Eq:pig}
\end{equation}
which, through Fourier transformation, gives
\begin{equation}
\tilde\Phi_{\pi_g}(\vec k, t ) \propto \left(e^{\dot\imath R k_y} - e^{-\dot\imath R k_y}\right) \tilde f(\vec k , t)\,.
\label{Eq:pigFT}
\end{equation}
The squared modulus of this wave function behaves as $\sin^2(R\,k_y)$ and thus exhibits zeros for $k_y = n \pi/R$. The positions of the dark fringes should thus be directly related to the internuclear distance $R$. These positions are indicated by dotted vertical gray lines in Fig.\,\ref{figure2}. The local minima in the calculated spectrum $S(k_y)$ are seen to agree fairly well with these predicted positions. The complete quantitative analysis gives, within the model assumptions, $R \simeq 4.92\,$\AA, with a deviation of less than $3\%$ from the initial input value $R = 4.8\,$\AA.

Given the role that lower-lying valence molecular orbitals play for CO$_2$ in intense laser fields \cite{HHG_CO2}, the averaged $S(k_y)$ distribution for the HOMO-2 is shown as a green dotted line in the upper panel of Fig.\,\ref{figure2}. Though a bonding $\sigma_u$ orbital, the HOMO-2 presents the same symmetry property as the $\pi_g$ HOMO with respect to the $\sigma_h$ mirror plane containing the carbon atom. Eqs.\,(\ref{Eq:pig}) and (\ref{Eq:pigFT}) are thus valid for the HOMO-2 also, and the photo-electron signal exhibits the same interference pattern for large $k_y$ values. In contrast, the bonding $\pi_u$ HOMO-1, being symmetric with respect to the $\sigma_h$ mirror plane, gives rise to fringes out-of-phase with respect to those associated with the HOMO and HOMO-2. The nodal structure of this orbital is reflected through
\begin{equation}
\Phi_{\pi_u}(\vec r, t ) \propto f(x,y-R,t) + f(x,y,t) + f(x,y+R,t)\,,
\end{equation}
which, through Fourier transformation, gives
\begin{equation}
\tilde\Phi_{\pi_u}(\vec k, t ) \propto \left(e^{\dot\imath R k_y} + 1 + e^{-\dot\imath R k_y}\right) \tilde f(\vec k , t)\,.
\end{equation}
The squared modulus of this wave function, which behaves as $[1+2\cos(R\,k_y)]^2$, is characterized by a regular succession of two peaks of different amplitudes. Their maxima, located at $k_y = n \pi/R$, and their relative amplitudes are reproduced in the full quantum calculation shown in the lower panel of Fig.\,\ref{figure2}. The total photo-electron signal, constituted by combined contributions from the HOMO, HOMO-1 and HOMO-2 which can be controlled by the field intensity, could be used to read accurately the CO bond length from the fringe pattern observed in the large $k_y$-range.

\begin{figure}[t!]
\begin{center}
\includegraphics[width=0.48\textwidth]{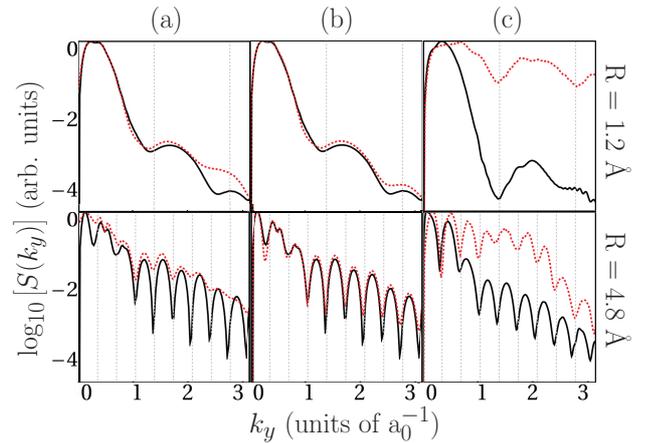}
\caption{\label{figure3} (Color online) Averaged normalized electron diffraction patterns $S(k_y)$ resulting from ionizing the HOMO of CO$_2$ under the same field conditions as in Fig.\,\ref{figure1} for $R = 1.2\,$\AA\ (first row) and $R = 4.8\,$\AA\ (second row). The solid black lines correspond to a perfect alignment at fixed internuclear distance for a single optical cycle pulse. The red dashed lines correspond to, column (a), an imperfect alignment, column (b), a distribution of $R$ and, column (c), a 10\,fs FWHM sine-square pulse.
}
\end{center}
\end{figure}

The structural reading made above of the diffraction pattern refers to a perfectly aligned molecule in a fixed nuclear geometry configuration. In reality, molecules are subject to rotations and vibrations, and these geometrical parameters are usually distributed over some range dictated by the rotational and vibrational state in which the parent molecule is initially prepared. A first aligning laser pulse can be used to create a rotational wave packet which periodically rephases, thus yielding, at the revival time, a strong molecular alignment along the polarization direction. Angular distributions confined in a cone angle of about 20$^\circ$ have been reached experimentally with CO$_2$\,\cite{CO2-align}. The left column of Fig.\,\ref{figure3} shows, in red dotted lines, the diffraction pattern incoherently averaged over a Gaussian distribution of alignment angles $\theta$, with a characteristic width of $20^{\circ}$ for $R = 1.2\,$\AA\ (first row) and $R = 4.8\,$\AA\ (second row). This Figure also shows, in black solid lines, the result obtained for a perfect alignment. The contrast loss seen in this Figure denotes a certain sensitivity to alignment defects. However, the fringes are still clearly observable at their original positions in the range $|k_y| > 0.7$\,a.u. The extraction of geometrical informations thus remains achievable even in case of imperfect initial alignment of the molecular target. To study the effect of variations of the bond length, we also considered a Gaussian distribution of internuclear distances with a characteristic width of 0.2\,\AA. Note that this width is larger than the typical size of the ground vibrational level associated with the stretching and bending modes of CO$_2$. The results are shown in the middle column of Fig.\,\ref{figure3}. One can notice that the diffraction pattern $S(k_y)$ is almost insensitive to the initial distribution of $R$. The zeroes in the averaged $S(k_y)$ remain basically unshifted and can therefore be used for geometrical information and imaging purposes. To conclude, we expect that a measurement on imperfectly aligned CO$_2$ molecules in their ground vibrational level would provide diffraction patterns that are no less analyzable than those shown in Fig.\,\ref{figure2}, and from which the CO equilibrium bond length could be extracted unambiguously. We also note in Fig.\,\ref{figure3} that, as expected, the oscillation period of $S(k_y)$ is 4 times smaller with $R=4.8\,$\AA\ than with $R=1.2\,$\AA.

Finally, the right column of Fig.\,\ref{figure3} shows the results obtained with a $10\,$fs FWHM sine-square pulse of same carrier wavelength, $800\,$nm. In this case, the 2D photo-electron momentum spectrum (not shown) is different from the one obtained under the single-cycle pulse (Fig.\,\ref{figure1}). It is more symmetric with respect to $k_x$, reflecting a different rescattering physics: Concentrating on what happens near the peak of the pulse envelope, one distinguishes two separate electron trajectories, which correspond to ionization events taking place at different times. The associated electron wave packets are then driven by the field in opposite directions. Contrary to what happens in the single-cycle case, these trajectories can both return to the ion and contribute to the diffraction event. The returning electron waves overlap and interfere with each other in the low-energy region of the 2D momentum distribution, corresponding to $|k| \le 2\,$a.u., blurring out the diffraction pattern and prohibiting a simple reading of the fringes in this region. However, even with a long pulse duration, only a few trajectories contribute to the highest energy domain of the spectrum. An average of the 2D diffraction map over the domain $\Gamma$ of Eq.\,(\ref{Eq:S}) corresponding for instance to $|k_x| > 3.15\,$a.u. helps recovering clearly visible interference fringes. In the right column of Fig.\,\ref{figure3}, the black solid line shows the result obtained using this integration domain in the case of a single cycle pulse, while the red dotted line refers to a $10\,$fs FWHM sine-square pulse. It is seen that with this pulse, not only does the diffraction signal $S(k_y)$ continue to reveal the geometrical and orbital structure of the parent molecule, but it also is amplified by about an order of magnitude as compared to the single-cycle signal.

To conclude, we have suggested a simple yet robust method to extract the molecular structure from the photo-electron spectra of a laser-driven linear, symmetric molecule. It is applicable to any sufficiently symmetric molecular system for which the HOMO, or any molecular orbital contributing to the ionization signal, has specific symmetry properties that would allow for a simple interpretation of the diffraction patterns. 
The detailed structure of the diffraction image reflects the symmetry of the molecular orbital from which the recolliding electron emanates since this symmetry is conserved during the time-evolution of the system under the combined effect of the Coulomb forces and the laser field. Such a symmetry conservation actually goes beyond the SAE approximation, and we expect the diffraction images calculated here to survive inclusion of many-electron effects, contrary to what was found relative to corrections to the strong field approximation\,\cite{Saenz}, which is not assumed here.

As the acquisition process, relying on the rescattering mechanism, is faster than nuclear motions, it can be used to produce a sequence of stroboscopic pictures, making up, for example, a movie of vibrational dynamics. The imaging technique could be implemented using electron spectroscopy based on different schemes such as angle-resolved time-of-flight detection or velocity map imaging \cite{Ghafur}. These methods are currently used in the atomic, molecular and optical physics community, and offer an energy resolution $\Delta\varepsilon/\varepsilon$ of the order or better than 1\% at 50\,eV, which turns out to be largely sufficient to resolve the fringe separations suggested by the present simulations. While, in the present theoretical study, a typical 800 nm laser radiation was considered, preliminary numerical simulations also show that our proposal could be used efficiently with few-cycle pulses produced nowadays in the IR domain (1-3\,$\mu$m) \cite{fcpIR}. Practically, a single-cycle pulse has been produced at $\lambda$=1.8 $\mu$m \cite{FLG}. Higher recollision energies are then obtained, yielding shorter electron de Broglie wavelengths. These laser systems should therefore provide a better resolution in terms of internuclear distances at fixed laser intensity. 

M.P. and T.T.N.D. acknowledge the Natural Sciences and Engineering Research Council of Canada (NSERC) for financial supports. The authors acknowledge supports from CFQCU (contract 2010-19), from ANR (contracts ImageFemto ANR-07-BLAN-0162 and Attowave ANR-09-BLAN-0031-01), from FCS Digiteo - Triangle de la Physique (project 2010-078T - High Rep Image), and from the EU (ITN-2010-264951 - CORINF).

\bibliographystyle{apsrev4-1}

\end{document}